\documentclass[iop]{emulateapj}

\usepackage{braket}

\shorttitle{}
\shortauthors{Hasegawa et al}

\begin{document}

\title{The Heavy-element Content Trend of Planets: A Tracer of their Formation Sites}

\author{Yasuhiro Hasegawa\altaffilmark{1}, Bradley M. S. Hansen\altaffilmark{2}, and Gautam Vasisht\altaffilmark{1}}
\affil{$^1$Jet Propulsion Laboratory, California Institute of Technology, Pasadena, CA 91109, USA}
\affil{$^2$Mani L. Bhaumik Institute for Theoretical Physics, Department of Physics \& Astronomy, University of California Los Angeles, Los Angeles, CA 90095, USA}

\email{yasuhiro.hasegawa@jpl.nasa.gov}

\begin{abstract}
Identification of the main planet formation site is fundamental to understanding how planets form and migrate to the current locations.
We consider the heavy-element content trend of observed exoplanets derived from improved measurements of mass and radius,
and explore how this trend can be used as a tracer of their formation sites.
Using gas accretion recipes obtained from detailed hydrodynamical simulations,
we confirm that the disk-limited gas accretion regime is most important for reproducing the heavy-element content trend.
Given that such a regime is specified by two characteristic masses of planets,
we compute these masses as a function of the distance ($r$) from the central star, and then examine how the regime appears in the mass-semimajor axis diagram.
Our results show that a plausible solid accretion region emerges at $r \simeq 0.6$ au and expands with increasing $r$, using the conventional disk model.
Given that exoplanets that possess the heavy-element content trend distribute currently near their central stars,
our results imply the importance of planetary migration that would occur after solid accretion onto planets might be nearly completed at $r \geq 0.6$ au. 
Self-consistent simulations would be needed to verify the predictions herein.
\end{abstract}


\section{Introduction} \label{sec:intro}

A remarkable feature revealed by observations is that exoplanetary systems exhibit great diversity \citep{wf15}.
The diversity has led to the proposition of a number of formation and migration scenarios.
These include pebble accretion for the efficient build-up of planetary cores at $r>10$ au \citep{ok10,lj12},
in-situ gas accretion for forming hot Jupiters at the present close-in locations \citep{bhl00,bbl16},
and planetary migration driven by disk-planet interaction \citep{lbr96,kn12}.
In addition to the standard core accretion scenario \citep{p96},
exploitation of these mechanisms allows the possibility of reproducing a wealth of exoplanets' observational properties \citep{il04i,mka14,h16,jl17}.
 
Despite the progress, our understanding of planet formation in protoplanetary disks is nevertheless imperfect.
One critical reason for this is that the primary formation site of planets is poorly constrained.
If the site could be identified, one can infer which formation mechanism(s) would dominate and to what extent, planetary migration would be needed 
for explaining the current orbital architecture of both the solar and extrasolar planetary systems.
It has been suggested recently that the carbon-to-oxygen (C/O) ratio of planets is one promising observable for identifying where planets form \citep{omb11,mak14,bfm17}.
However, \citet{mvm16} point out that an improved understanding of the distribution of elemental materials in natal protoplanetary disks would be needed
in order to derive any useful information from the observed C/O ratio \citep[also see][]{efm17}.

Here we propose another quantity as a tracer for the formation site of exoplanets.
Improved measurements of masses and radii of exoplanets enable the computation of the abundance of heavy elements in a well-measured subset of planets \citep{gsp06,mf11}.
Through a careful selection of 47 exoplanets taken from larger samples, 
\citet[hereafter, T16]{tfm16} derive the following correlations between a planet's total mass ($M_p$) and its heavy-element mass ($M_Z$)
and between $M_p$ and its metallicity ($Z_p \equiv M_Z/M_p$). These are, respectively
\begin{equation}
\label{eq:T16}
M_Z \propto M_p^{\Gamma_Z} \mbox{ and } \frac{Z_p}{Z_s} \propto M_p^{\beta_Z},
\end{equation} 
where $Z_s$ is the host stellar metallicity, $\Gamma_Z=0.61 \pm 0.08 $, and $\beta_Z=-0.45 \pm 0.09$.
These correlations are referred to as the heavy-element content trend in this work.
The follow-up work of \citet[hereafter H18]{hbi18} provides an explanation for this trend, 
focusing on solid accretion from {\it gapped} planetesimal disks. 
Such solid accretion occurs simultaneously with gas accretion after planetary core formation is completed.
In this Letter, we use the heavy-element content trend and the analysis of H18 
to identify a plausible solid accretion zone in the mass-semimajor axis diagram.
Our study implies that in order to reproduce the trend, 
planets would initially, efficiently accrete solids beyond $r \geq 0.6$ au and subsequently migrate to their present locations.

\section{Metal Enrichment of Planets via Planetesimal accretion} \label{sec:mod}

\subsection{Disk model} \label{sec:mod_1}

We adopt the steady state disk model \citep{fkr02}:
\begin{equation}
\label{eq:mdot}
\dot{M}_{\rm d} = 3 \pi \nu \Sigma_{\rm g},
\end{equation}
where $\dot{M}_{\rm d}$ is the disk accretion rate, $\Sigma_{\rm g}$ is the gas surface density,
$\nu = \alpha c_{\rm s} H_{\rm g}$  is the effective viscosity, $c_{\rm s}$ is the local sound speed, $H_{\rm g}=c_{\rm s}/ \Omega$ is the pressure scale height, 
and $\Omega$ is the Keplerian angular velocity.
The $\alpha$-prescription is used for characterizing the efficiency of angular momentum transport in disks \citep{ss73}.
For the disk temperature ($T_{\rm d}$) prescription, we follow the minimum-mass solar nebula model \citep{h81}:
\begin{equation}
\label{eq:T_d}
T_{\rm d} = T_{\rm d0} \left( \frac{r}{1 \mbox{ au}} \right)^{-t},
\end{equation}
where $T_{\rm d0}=280$ K and $t=1/2$ under the assumption of an optically thin disk.

There are two parameters in this disk model, $\dot{M}_{\rm d}$ and $\alpha$.
We verify that results of this work are relatively insensitive to variations in $\dot{M}_{\rm d}$; therefore we adopt $\dot{M}_{\rm d}= 10^{-8} M_{\odot} \mbox{ yr}^{-1}$,
following disk observations \citep{hcg98,wc11}.
We assume that $\alpha=10^{-2}$, a choice is motivated by the recent MHD simulations of protoplanetary disks.
When disks are fully ionized and non-ideal MHD effects are of lesser importance,
the magnetorotational instability operates and the resulting MHD turbulence transports angular momentum radially \citep{bh98}.
When non-ideal MHD effects dominate and magnetic fields threading disks are strong enough, and with appropriate geometries,
disks would be laminar and magnetically induced disk winds remove angular momentum vertically \citep{si09,bs13}.
In both cases, the corresponding value of $\alpha$ is the order of $10^{-2}-10^{-3}$ to account for high accretion rates \citep{hcg98,oh11,hof17}.

\subsection{Gas and solid accretion onto planets} \label{sec:mod_2}

We consider the metal enrichment of planets through planetesimal accretion, after core formation is complete.
In this case, the efficiently of planetesimal accretion is related to the rates of gas accretion onto (proto)planets \citep[H18]{zl07,si08}.
We now describe the model used in this work.

First, we consider solid accretion, for which we employ a semi-analytical approach.
It would be ideal to compute the total heavy-element mass accreted onto planets by tracing planet formation and migration histories.
When a large parameter space would be covered by running population synthesis calculations,
one can directly compare theoretical predictions with observational results \citep{mka14}.
In this work, however, we focus exclusively on the power-law indices ($\Gamma_Z$ and $\beta_Z$) for the heavy-element content trend.
This is because then one can examine each planet-forming process individually, and 
specify what process would be most crucial for understanding the inferred trend.
In practice, we closely follow the approach in H18,
wherein a semi-analytical formula for planetesimal accretion rates, derived from detailed $N$-body simulations \citep{si08}, is employed.
Assuming that the planet radius scales as $M_p^{1/3}$, $\Gamma_Z$ and $\beta_Z$ can be written as (H18)
\begin{eqnarray}
\label{eq:Mpz}
\Gamma_{Z} & = &  1 + \beta_{Z}  \\ \nonumber
                      & = & 1 - \frac{12D +17}{30}  = \frac{13 - 12D }{30}, 
\end{eqnarray}
where $D$ is the power-law index of the gas accretion timescale, that is, $\tau_p = M_p / \dot{M}_p \propto M_p^D$,
where $\dot{M}_p$ is the gas accretion rate onto planets.
By definition, $Z_p \equiv M_Z/M_p$, that is, $\Gamma_{Z}  =  1 + \beta_{Z}$.
Note that the above equation is derived under the assumption 
that solid accretion onto planets takes place from gapped planetesimal disks without migration (see below for the importance of planetesimal gaps).
Thus, one can compute the values of $\Gamma_Z$ and $\beta_Z$ directly for given values of gas accretion rates via $D$.

Gas accretion onto planets becomes possible when the surface escape velocity of (proto)planets exceeds the sound speed of the surrounding disk gas.
This corresponds roughly to moon-mass objects at $r=1$ au in our model.
Such accreted gas forms hydrostatic envelopes around planetary cores due to the pressure gradient.
Gas accretion contributes effectively to planetary growth when the hydrostatic assumption breaks down and the envelopes contract rapidly \citep{p96}.
The critical core mass is defined for this transition, 
and the gas accretion rate is initially determined by the timescale of envelope contraction, also known as the Kelvin-Helmholtz timescale \citep{ine00}:
\begin{equation}
\label{eq:tau_KH}
\tau_{p, \rm KH} = 10^{c} f_{\rm grain} \left( \frac{M_p}{ 10 M_{\oplus} } \right)^{-d} \mbox{yr},
\end{equation}
where $f_{\rm grain} \ll 1$ is the acceleration factor due to the reduction of grain opacity in planetary envelopes,
and we set that $c=7$ and $d=4$ following \citet{tn97}.
Recent studies suggest that dust growth and sedimentation are efficient in planetary envelopes \citep{mbp10,o14}
and that the resulting reduction in grain opacity is preferred for better reproducing the population of observed exoplanets \citep{mka14,hp14}.
We therefore assume that $f_{\rm grain} = 10^{-2}$.
Consequently, the mass growth rate ($\dot{M}_{p, \rm KH}$) of planets is written as
\begin{equation}
\label{eq:Mpdot_KH}
\dot{M}_{p, \rm KH}  \simeq         \frac{M_p}{ \tau_{p, \rm KH} } 
                                = 10^{-4}  \left( \frac{f_{\rm grain}}{ 10^{-2} } \right)^{-1}  \left( \frac{M_p}{ 10 M_{\oplus} } \right)^{5} \frac{M_{\oplus}}{\mbox{yr}}.
\end{equation}

The value of $\dot{M}_{p, \rm KH}$ increases rapidly  with increasing $M_p$.
In order to avoid an unrealistically high value of $\dot{M}_{p, \rm KH}$, 
we use the results of hydrodynamical simulations \citep{tw02} and impose the following upper limit \citep{ti07}:
\begin{eqnarray}
\label{eq:Mpdot_hydro}
\dot{M}_{p, \rm hydro} & =          &  0.29 \left( \frac{H_{\rm g}}{r_p} \right)^{-2}  \left( \frac{M_p}{M_*} \right)^{4/3} \Sigma_{\rm g} r_p^2 \Omega  \\ \nonumber
                                   & \simeq &  1.5 \times 10^{-3}  \left( \frac{\alpha}{10^{-2}} \right)^{-1} \left( \frac{H_{\rm g}/r_p}{0.05} \right)^{-4}  \\ \nonumber
                                   & \times  &   \left( \frac{M_p}{10 M_{\oplus}} \right)^{4/3} \left( \frac{\dot{M}_{\rm d}}{10^{-8} M_{\odot} \mbox{ yr}^{-1}}  \right)       \frac{M_{\oplus}}{\mbox{yr}},
\end{eqnarray}
where $r_p$ is the position of planets and $M_*=M_{\odot}$ is the mass of the central star.
Thus, as the planet mass increases,  gas supply from disks to planets is limited by disk evolution.

The above expressions will remain valid until planets are massive enough to open up gaps in gas disks \citep{kn12}.
Once planet-disk interaction starts modifying the disk structure, the gas accretion flow will come from the polar direction rather than along the midplane \citep{mki10,smc14}.
Assuming that the gas dynamical timescale is $\tau_{\rm dyn} \sim H_{\rm g}^2 / \nu$ and that the gas accretion flow originates from $z \geq r_{\rm H}$,
where $r_{\rm H}= r_p(M_p/(3M_*))^{1/3}$ is the Hill radius of planets, 
the gas accretion rate onto planets ($\dot{M}_{p, \rm gapI}$) can be given as \citep{msc14}
\begin{eqnarray}
\label{eq:Mpdot_gapI}
\dot{M}_{p, \rm gapI} & \simeq         & 2 \pi r_p v_r 4 
                                                   \int_{r_{\rm H}}^{\infty} dz \frac{\Sigma_{\rm g}}{ \sqrt{ 2 \pi }H_{\rm g} } \exp \left( - \frac{z^2}{2 H_{\rm g}^2} \right)  \\ \nonumber
                                  & =         &   \frac{4}{3}  \left( \frac{H_{\rm g}}{r_p} \right)^{-1}   
                                                      \mbox{erfc} \left[ \frac{1}{ \sqrt{2}} \left( \frac{H_{\rm g}}{r_p} \right)^{-1} \left( \frac{M_p}{3 M_*} \right)^{1/3} \right]   \dot{M}_{\rm d},
\end{eqnarray}
where $v_r = H_{\rm g} / \tau_{\rm dyn}$ is the gas radial velocity and $\mbox{erfc}$ is the complementary error function.
A factor of 4 arises to take account of the accretion flow coming from two surface layers of the disk and both sides of a gas gap.

There are other gas accretion recipes available in the literature \citep{lhdb09,tt16}.
As an example, we consider the one ($\dot{M}_{p, \rm gapII} $) that utilizes the results of more recent hydrodynamical simulations \citep{fsc14,kmt15}.
These simulations suggest that gas gaps carved by planets tend to be shallower than those predicted by previous simulations \citep{ti07,lhdb09}.
Then the resulting $\dot{M}_{p, \rm gapII} $ is given as \citep{tt16}
\begin{eqnarray}
\label{eq:Mpdot_gapII}
\dot{M}_{p, \rm gapII} & =          & \frac{8.5}{3 \pi} \left( \frac{H_{\rm g}}{r_p} \right) \left( \frac{M_p}{M_*} \right)^{-2/3}  \dot{M}_{\rm d} \\ \nonumber
                                 & \simeq & 3.4 \times 10^{-2}  \left( \frac{H_{\rm g}/r_p}{0.05} \right) \left( \frac{M_p}{100M_{\oplus}} \right)^{-2/3}  \\ \nonumber
                                 & \times  &   \left( \frac{\dot{M}_{\rm d}}{10^{-8} M_{\odot} \mbox{ yr}^{-1}}  \right)  \frac{M_{\oplus}}{\mbox{yr}}.
\end{eqnarray}
In our preliminary study, we have found that as the planet mass increases,
our recipe ($\dot{M}_{p, \rm gapI}$) takes the intermediate value between $\dot{M}_{p, \rm gapII}$ and the one derived from the classical deep gap.
Thus, $\dot{M}_{p, \rm gapI}$ provides the mean behavior of gas accretion onto planets after gas gap formation. 

In summary, we consider 4 gas accretion recipes to compute the values of $\Gamma_Z$ and $\beta_Z$ (see Equation (\ref{eq:Mpz})).

\subsection{Resulting trends of the heavy-element mass} \label{sec:mod_3}

We discuss what stage of planet formation is most important for reproducing the results of T16 (see Equation (\ref{eq:T16})).

\begin{figure*}
\begin{minipage}{17cm}
\begin{center}
\includegraphics[width=8cm]{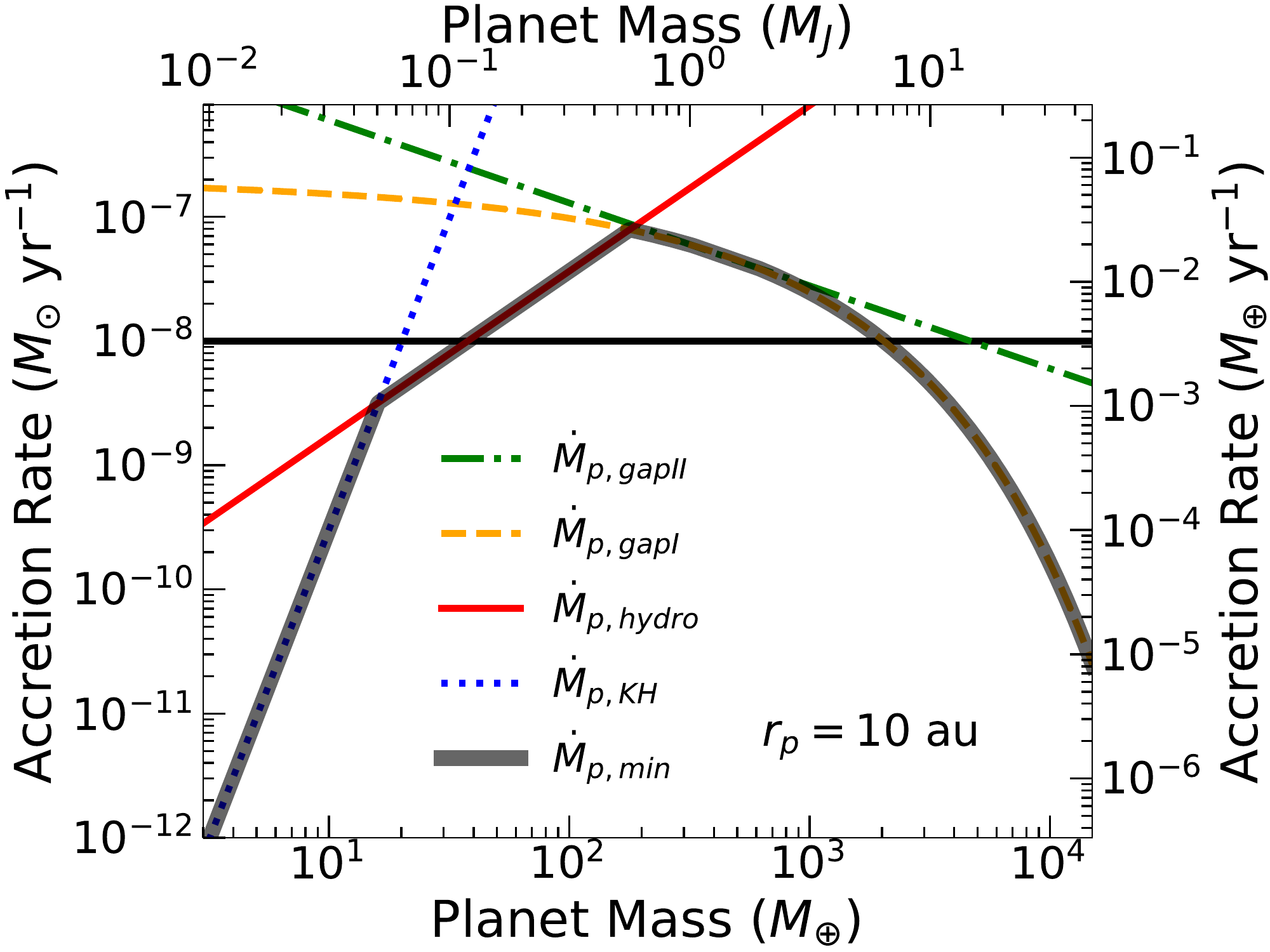}
\includegraphics[width=8cm]{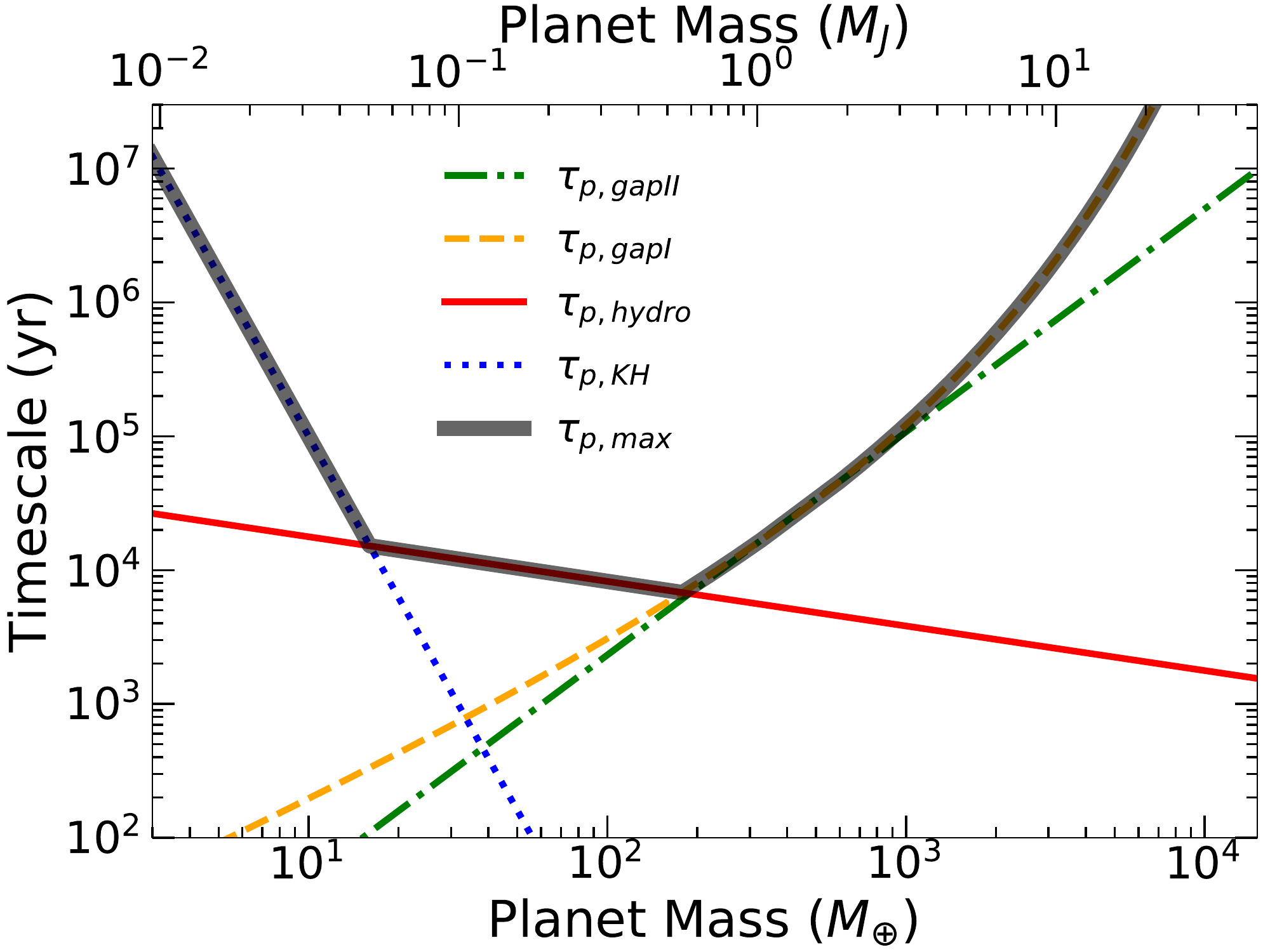}
\caption{Gas accretion rates and the resulting timescales as a function of planet mass on the left and right panels, respectively.
The case that $r_p=10$ au is considered here.
On the left, $\dot{M}_{\rm d}$ is denoted by the horizontal, black line for reference.
The minimum value of $\dot{M}_p$ is denoted by the thick gray line.
On the right, gas accretion timescales are computed by $\tau_{p}=M_p/ \dot{M}_p$.}
\label{fig1}
\end{center}
\end{minipage}
\end{figure*}
                       
Figure \ref{fig1} shows gas accretion rates and the resulting timescales as a function of planet mass on the left an right panels, respectively.
We consider that $r_p=10$ au here.
Gas accretion switches from $\dot{M}_{p, \rm KH}$ to $\dot{M}_{p, \rm hydro}$, and to the one ($\dot{M}_{p, \rm gapI}$ or $\dot{M}_{p, \rm gapII}$) with increasing planet mass.
We find that while the recipe of $\dot{M}_{p, \rm gapI}$ is rather simple,
the resulting value becomes comparable to that of $\dot{M}_{p, \rm gapII}$ when planets become just massive enough to open up gaps in the disks.
As the planet mass increases, $\dot{M}_{p, \rm gapI}$  becomes smaller than $\dot{M}_{p, \rm gapII}$
since the latter can achieve efficient gas accretion due to shallower gaps.
We focus on computing the values of $\Gamma_Z$ and $\beta_Z$,
and hence it is assumed in Equations (\ref{eq:Mpdot_gapI}) and (\ref{eq:Mpdot_gapII}) that 
planets accrete the disk gas flowing into gaps at the 100 \% efficiency.
Numerical simulations, however, show that only some fractions of gas can contribute to planetary growth and 
the rest of gas goes back to the surrounding disks through the horse-shoe orbit \citep{ld06}.
This example exhibits that the first ($\dot{M}_{p, \rm KH}$) and the final ($\dot{M}_{p, \rm gapI}$ or $\dot{M}_{p, \rm gapII}$) stages take a longer time
and planets can become gas giants within gas disk lifetimes when their core mass is $\ga 5 M_{\oplus}$.

\begin{figure}
\begin{center}
\includegraphics[width=8cm]{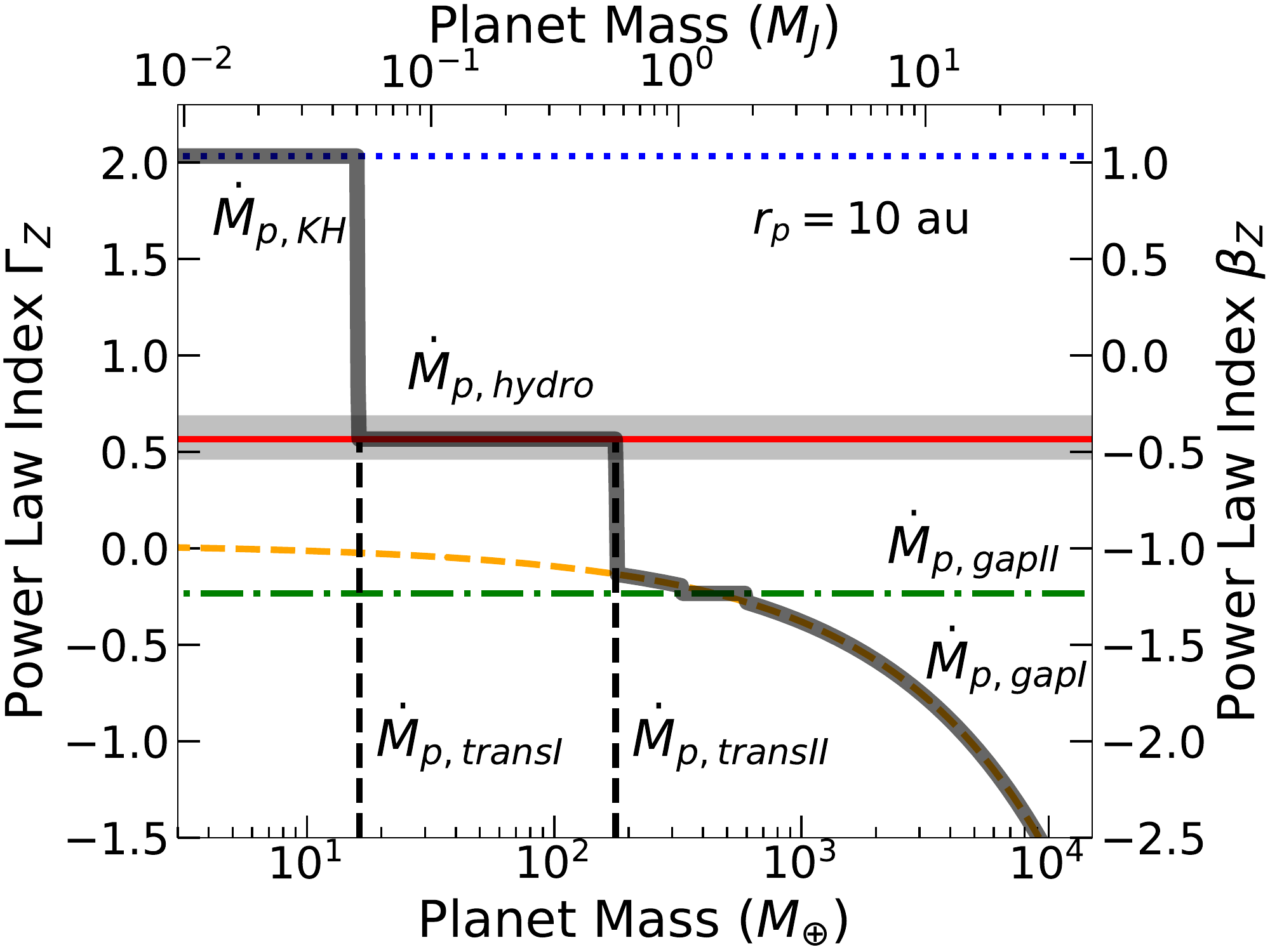}
\caption{The resulting power-law indices ($\Gamma_Z$ and $\beta_Z$) as a function of planet mass 
for the case that solid accretion onto planets occurs from gapped planetesimal disks.
The thick gray line denotes the results obtained from  $\dot{M}_{p, \rm min}$ (see Figure \ref{fig1}).
The horizontal, shaded silver region represents the results of T16 that can be reproduced by the regime of $\dot{M}_{p, \rm hydro}$ 
(see the vertical dashed lines marked by $M_{p, \rm transI}$ and $M_{p, \rm transII}$).}
\label{fig2}
\end{center}
\end{figure}
 
We now compute the values of $\Gamma_Z$ and $\beta_Z$ by adopting $D \, (M_p / \dot{M}_p \propto M_p^D)$ obtained from each gas accretion regime.
Figure \ref{fig2} shows the results.
The values of $\Gamma_Z$ and $\beta_Z$ change suddenly when the gas accretion recipe switches from one another as the planet mass increases (see the thick gray line).
We confirm the finding of H18:
The heavy-element content trend can be reproduced well if it traces the stage where
planets accrete solids from their surrounding, {\it gapped} planetesimal disks, while gas accretion is limited by disk evolution.
In other words, there is a plausible mass range for explaining the heavy-element content trend 
(see the regime of $\dot{M}_{p, \rm hydro}$ encompassed by $M_{p, \rm transI}$ and $M_{p, \rm transII}$).
We examine below how such a mass range behaves as a function of the distance from the central star.

\subsection{Formation sites} \label{sec:mod_4}

We finally identify formation sites of planets that can reproduce the heavy-element content trend.

The plausible mass range is computed by $\dot{M}_{p, \rm KH}=\dot{M}_{p, \rm hydro}$ and $\dot{M}_{p, \rm hydro}=\dot{M}_{p, \rm gapII}$ (see Figure \ref{fig2}).
Physically, the range represents planets that are more massive than the critical core mass and undergo efficient gas accretion (the former condition) 
and that are less massive to open up gaps in gas disks (the latter).
Then the characteristic masses can be written as, respectively
\begin{eqnarray}
\label{eq:Mp_transI}
M_{p, \rm transI} & \simeq & 21 \left( \frac{f_{\rm grain}}{ 10^{-2} } \right)^{3/11}   \left( \frac{\alpha}{10^{-2}} \right)^{-3/11}  \\ \nonumber
                            & \times &\left( \frac{H_{\rm g}/r_p}{0.05} \right)^{-12/11}  \left( \frac{\dot{M}_{\rm d}}{10^{-8} M_{\odot} \mbox{ yr}^{-1}}  \right)^{3/11}  M_{\oplus},
\end{eqnarray}
\begin{equation}
M_{p, \rm transII} \simeq  0.3 \left( \frac{\alpha}{10^{-2}} \right)^{1/2} \left( \frac{H_{\rm g}/r_p}{0.05} \right)^{5/2} M_{\rm J}, 
\end{equation}
where $M_{\rm J}$ is Jupiter mass.
Note that in order to compute $M_{p, \rm transII}$, we have adopted $\dot{M}_{p, \rm gapII}$, rather than $\dot{M}_{p, \rm gapI}$.
This substitute simplifies the expression of $M_{p, \rm transII}$ considerably 
without any significant deviation from the mass computed by $\dot{M}_{p, \rm hydro}=\dot{M}_{p, \rm gapI}$.

\begin{figure}
\begin{center}
\includegraphics[width=8cm]{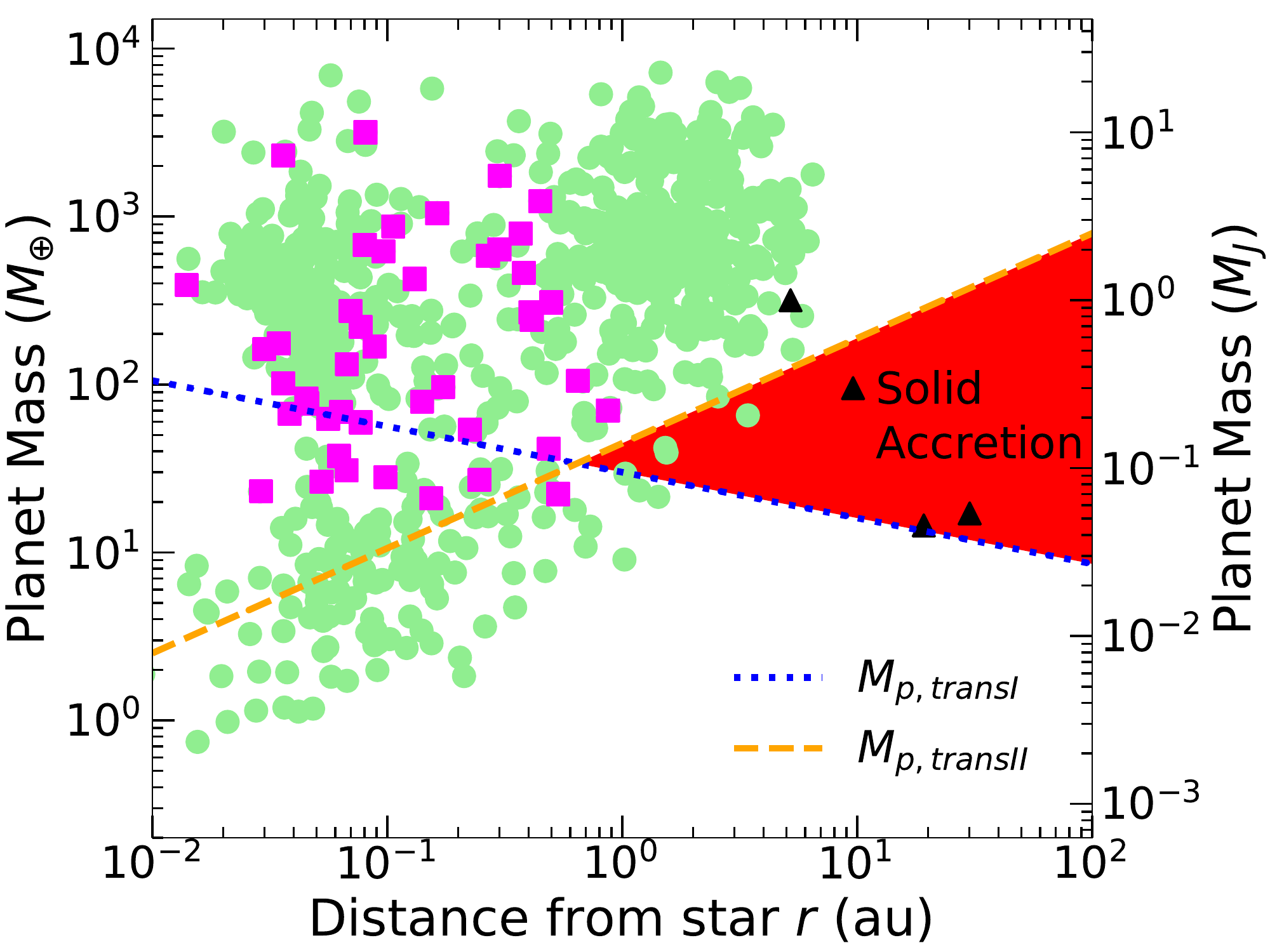}
\caption{The radial dependences of $M_{p, \rm transI}$ and $M_{p, \rm transII}$, and 
the resulting mass region in which the heavy-element content trend can be reproduced (see the red shaded region).
The light green dots denote observed exoplanets listed in exoplanets.org, while the purple squares are the samples of T16.
For reference,  the solar system planets are shown by the black triangles.}
\label{fig3}
\end{center}
\end{figure}

Figure \ref{fig3} shows the resulting values of $M_{p, \rm transI}$ and $M_{p, \rm transII}$ and the corresponding mass region.
Our results indicate that the plausible mass region emerges from $r \simeq 0.6$ au and expands with increasing $r$.
This is the direct reflection of the radial dependences of $M_{p, \rm transI}$ and $M_{p, \rm transII}$.
The former is a decreasing function of $r$ due to $\dot{M}_{p, \rm hydro}$.
In the disk-limited gas accretion regime, shock around planets regulates the gas accretion efficiency \citep{tw02}.
At a larger value of $r_p$, the disk temperature decreases and less massive planets can excite shock readily.
On the other hand, $M_{p, \rm transII}$ increases with increasing $r$.
This occurs simply because gas gap formation due to disk-planet interaction becomes less efficient as $r$ increases.
Thus, our analysis shows that the heavy-element content trend can be reproduced well
if planets accrete solids efficiently from gapped planetesimal disks at $r_p \geq 0.6$ au.
Generally, the outer region of disks is the preferred site of forming more massive planets that accrete more planetesimals.

\section{Discussion} \label{sec:disc}

We first discuss the most crucial implication of this work.
Our results suggest that planets should undergo solid accretion efficiently at $r \geq 0.6$ au, in order to reproduce the heavy-element content trend (Figure \ref{fig3}).
It is however interesting that exoplanet samples employed in T16 currently distribute at much smaller $r$.
This implies that these exoplanets would have initially formed in the outer disks and subsequently migrated to the current locations.
It is beyond the scope of this work to explore which mode of migration was dominant for these exoplanets (either gas-induced migration or the one originating from $N$-body dynamics).
Nonetheless, our calculations demonstrate clearly that 
the heavy-element content trend can be used as an indicator of the importance of migration for understanding the present orbital architecture of (exo)planets.

Another implication obtained from our work may be for the C/O ratio of exoplanets' atmospheres.
As described in Section \ref{sec:intro},
its role is unclear currently due to lack of the knowledge about the spatial distribution of chemical materials in disks.
If observed exoplanets would follow the heavy-element content trend and their C/O ratios would be available,
then our results would play a complementary role in providing a tighter constraint on the primordial distribution of the C/O ratio in protoplanetary disks.

Our study bases on a number of assumptions and the mixture of semi-analytical formulae derived from numerical simulations that were run independently.
In particular, the most uncertain assumption in our model is that the planet radius is given by $M_p^{1/3}$ at all the stages of gas accretion.
This is motivated by the results of \citet{tn97} which show that planetary envelopes contract quasi-hydrostatically from Neptune-mass planets even up to Jovian mass planets.
It would, nonetheless, be natural to expect that dynamical collapse would occur at a certain time when planets become massive enough.
Then, planet radius would be expressed by a different functional form of $M_p$.
Assuming that $R_p \propto M_{p}^{q}$, the resulting values of $\Gamma_Z$ and $\beta_Z$ will shift by $2(q-1/3)$ (H18).
Another uncertainty in this work is that while the importance of planetary migration is suggested in Figure \ref{fig3},
solid accretion that can occur during migration is not considered.
Depending on the speed and timing of migration, planets can accrete solid at that time \citep{ti99}.
If its amount would be large enough, the heavy-element content trend generated at $\ga 0.6$ au will be washed out
and deviate from the linear correlation with the disk metallicity.
Formation of nearby planets would also cause similar effects by scattering planetesimals into the feeding zone of planets (see Section 5.2 of H18).
We have explored the parameter dependence on our results and found that the variation of $\alpha$ provides the largest change:
the plausible accretion zone emerges at $r \simeq 5$ au when $\alpha=10^{-3}$.
The zone shrinks somewhat when $T_{\rm d0}$ decreases and $t$ increases (Equation (\ref{eq:T_d})).
Verification of our work is thus demanded by running detailed simulations in a consistent and unified manner.

We should point out that if planetary cores are very massive ($> 20 M_{\oplus}$) and their cores dissolve into the envelopes 
as suggested for Jupiter \citep{whm17},
additional solid accretion might not be needed.
However, this scenario would work only for less massive ($\la 100 M_{\oplus}$) planets,
given that $M_Z$ is much larger than $ 20 M_{\oplus}$ for most giant planets examined in T16.

Finally, it is currently not clear how the heavy-element content trend is universal for all the observed, massive exoplanets.
If it would be the case, massive planets should accrete most of heavy elements in the plausible region, accompanying with efficient gas accretion.
Note that gas accretion with inefficient solid accretion might continue beyond the region, which corresponds to the stage after gas gap formation.
More and better measurements of mass and radius of exoplanets will answer this question.

In the near future, more exoplanet observations and better modeling of planet formation would be available 
not only for drawing a better picture of how and where planets accrete gas and solid from protoplanetary disks,
but also for testing our model of planet formation.

\acknowledgments

The authors thank an anonymous referee for useful comments on our manuscript.
This research was carried out at JPL/Caltech, under a contract with NASA.
Y.H. is supported by JPL/Caltech.

\bibliographystyle{apj}          


\end{document}